\RequirePackage{fixltx2e} 
\RequirePackage{xparse}
\documentclass[aip,reprint,onecolumn,groupedaddress]{revtex4-1}

\usepackage{amsmath,amssymb,amsfonts,dsfont}
\usepackage{graphicx,float}
\usepackage{braket}
\usepackage{cleveref}
\usepackage{color}
\usepackage{soul} 
\usepackage{siunitx}
\usepackage{booktabs}
\usepackage{subcaption}
\usepackage[justification=centerfirst,format=hang,singlelinecheck=0,font={sf,small}]{subfig}
\usepackage{setspace}
\usepackage{epstopdf}

\DeclareDocumentCommand\Diff{m}{\!\mathrm{d}#1\,}
\newcommand{\etal}{\textit{et al}. }
\newcommand{\comment}[1]{}

\newcommand{\chie}{\chi^{(3)}}
\newcommand{\chit}{\chi^{(2)}}

\newcommand{\ochie}{\overline{\chi}^{(3)}}
\newcommand{\nbar}{\overline{n}}
\newcommand{\me}{\mathrm{e}}
\newcommand{\vac}{\ket{\text{vac}}}

\DeclareDocumentCommand\dop{mm}{d^{#1}_{#2}}
\DeclareDocumentCommand\fop{mm}{f^{#1}_{#2}}
\DeclareDocumentCommand\aop{m}{\hat{a}_{#1}}
\DeclareDocumentCommand\bop{m}{\hat{b}_{#1}}

\DeclareDocumentCommand\aopd{m}{\hat{a}_{#1}^{\dagger}}

\begin{document}


\title{Proposal for an Integrated Raman-free Correlated Photon Source} 

\author{Daniel R. Blay}
\email[]{daniel.blay@mq.edu.au}
\author{L.~G. Helt}
\author{M.~J. Steel}
\affiliation{Macquarie University Quantum Research Centre in Science and Technology (QSciTech) and Centre for Ultrahigh bandwidth Devices for Optical Systems (CUDOS), MQ Photonics Research Centre, Department of Physics and Astronomy, Macquarie University, NSW 2109, Australia}

\date{\today}

\begin{abstract}
We propose a dual-pump third-order nonlinear scheme for producing pairs of correlated photons that is less susceptible to Raman noise than typical spontaneous four wave mixing methods (SFWM). Beginning with the full multimode Hamiltonian we derive a general expression for the joint spectral amplitude, from which the probability of producing a pair of photons can be calculated. As an example, we demonstrate that a probability of $0.028$ pairs per pulse can be achieved in an appropriately designed fused silica microfiber. As compared with single pump SFWM in standard fiber, we calculate that our process shows significant suppression of the spontaneous Raman scattering and an improvement in the signal to noise ratio.
\end{abstract}


\maketitle 
\singlespacing

\noindent The on-demand generation of single photons is keenly sought in quantum optics. There are a multitude of photon generation schemes, including atom-like sources \cite{Somaschi2016,Darquie2005,Hennessy2007,Albrecht2014}, and heralded photon pair sources based on nonlinear optics. Of the latter, the two most common schemes are spontaneous parametric down conversion (SPDC)\cite{Kwiat1995,Bonfrate1999, Tanzilli2001} and spontaneous four wave mixing (SFWM)\cite{Fiorentino2002,Inoue2004,Sharping2006}.

Due to the strength of the $\chit$ nonlinearity as compared with $\chie$, SPDC sources typically require lower pump powers than SFWM sources, and consequently
exhibit negligible noise from competing $\chie$ processes. However, most optical materials do not possess the symmetries required to have a $\chit$ response. Utilising the universal $\chie$ response allows for a greater number of materials and platforms to be used in heralded single photon sources, and integrates well with current telecommunication networks. In particular, silica and silicon allow for near infra-red generation and thus efficient coupling to standard SMF-28 fiber\cite{Fiorentino2002,Inoue2004}.

\begin{figure}
	\centering
	\begin{subfigure}{0.5\linewidth}
	\centering
	\includegraphics[width=.9\linewidth]{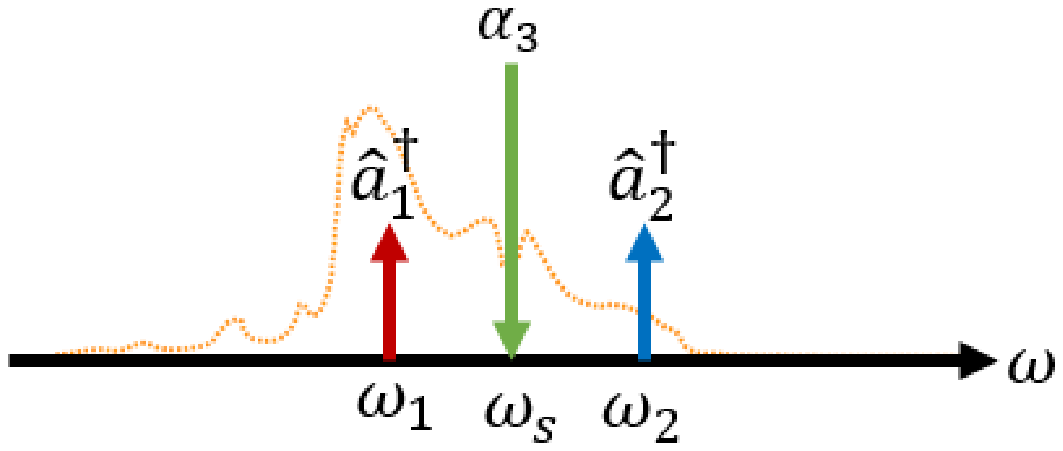}
	\caption{\label{subfig:sfwm_diagram}}
	\end{subfigure}%
	\begin{subfigure}{0.5\linewidth}
	\centering
	\includegraphics[width=.9\linewidth]{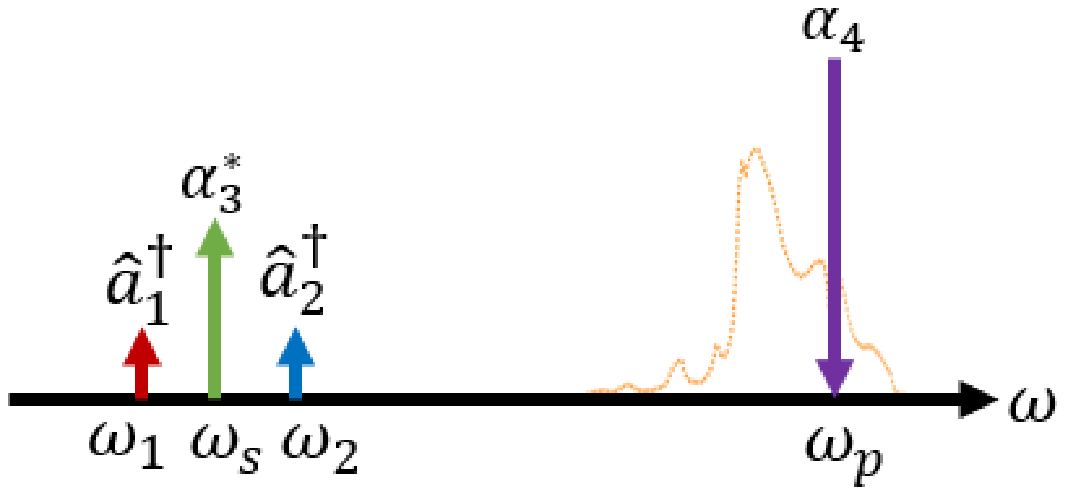}
	\caption{\label{subfig:sstpdc_raman}}
	\end{subfigure}
	\caption{Diagrammatic representations of the frequency channels involved in (a) degenerate SFWM, including the contamination of the produced pair from SpRS, and (b) SSTPDC, showing the pair generation spectrally distinct from the strong SpRS of the pump. The direction of the arrows denote photon generation or annihilation, and their length is suggestive of the powers involved, but is not to scale.}
\end{figure}

Many of these $\chie$ materials, such as fused silica, are amorphous and therefore exhibit broadband spontaneous Raman scattering (SpRS) due to inhomogenous broadening of the Raman transitions. In the quantum regime, this corresponds to emission of uncorrelated single photons. In a typical amorphous degenerate SFWM source, the strong pump field produces these uncorrelated Raman photons over a broad energy range.  This noise often overlaps with the desired frequency range of the generated pairs, \cite{Lin2007} (see \cref{subfig:sfwm_diagram}). Without due care, one may generate many more Raman photons than correlated pairs \cite{Collins2012a}.

Attempts have been made to mitigate Raman noise in amorphous SFWM sources \cite{Clark2011}, including dispersion engineering waveguides such that the produced pairs lie in a window of low Raman photon production \cite{Collins2012a,Li2006,He2012}. However, these are subject to material and engineering constraints that may be challenging to implement.

Here we propose a new method which sidesteps the issue of SpRS in amorphous materials. Pumping a $\chie$ material strongly at or near the third harmonic, $\omega_p \approx 3\omega_s$, three photons can be spontaneously generated at the third sub-harmonic $\omega_s$, which we call the fundamental, with low probability. When unstimulated this process is inefficient \cite{Bencheikh2007}, and so authors seeking three photon generation tend to use cascaded $\chit$ processes \cite{Hubel2010,Dot2012,Hamel2014}. As we are aiming for pair generation, here we  add a weak coherent field near the fundamental frequency $\omega_s$ to seed the process. This leaves the desired photon pair accessible, with energy conservation dictated by $\omega_p = \omega_1 + \omega_2 + \omega_s$. For a seed exactly at the fundamental, the pair of photons lie on either side of the seed, as seen in \cref{subfig:sstpdc_raman}. We may consider this process ``stimulated spontaneous three photon down conversion'' (SSTPDC). The principal advantage of this scheme is that the Stokes band of the Raman spectrum lies between the generated pairs and the pump frequency. The bandwidth of the spontaneous Raman response is typically of order 10~THz, whereas the fundamental and third harmonic fields involved at optical frequencies are separated by around 400~THz. This large spectral separation ensures low Raman noise in the signal band.

To describe this process and all of the multimode physics involved, we follow the formalism outlined by Yang, Liscidini and Sipe \cite{Yang2008}. We look for solutions to the first order Schr\"{o}dinger equation
\begin{equation}\label{eq:first_order_schro}
\ket{\psi_\text{out}} \approx \left[1 - \frac{i}{\hbar} \int_{t_0}^{t_1} \Diff{t} \hat{H}_\text{I}(t)\right] \ket{\psi_\text{in}},
\end{equation}
where $\ket{\psi_\text{in}}$ describes the coherent state input in the pump and seed modes, and is vacuum in the photon pair bands. As the nonlinearity outside the interaction length is zero, we are free to extend the integration limits to infinity, $t_0 \to -\infty$ and $t_1 \to \infty$. The relevant interaction Hamiltonian is given by
\begin{equation}\label{eq:full_3rdo_hamil}
\hat{H}_\text{I}(t)= -\frac{1}{4\epsilon_0}\int \Diff{^3\mathbf{r}} \Gamma^{(3)}_{ijmn} \hat{D}^i(\mathbf{r},t) \hat{D}^j(\mathbf{r},t) \hat{D}^m(\mathbf{r},t) \hat{D}^n(\mathbf{r},t).
\end{equation}
Here $\hat{D}^{(i)}(\mathbf{r},t)$ are the components of the vector displacement operator,  written in the 
interaction picture associated with the linear Hamiltonian ${\hat{H}_L = \sum_{\gamma} \int \Diff{k} \hbar \omega_{\gamma,k} \aopd{\gamma,k} \aop{\gamma,k}}$, (the summation is over the different modes $\gamma$, and the vacuum term is disregarded).
The symbol $\Gamma^{(3)}$ is a rank four tensor, related to the standard third order susceptibility tensor\cite{Yang2008} (see below).

To treat the SSTPDC process, we express the displacement field operator as a sum over modes $\gamma$,
\begin{align}\label{eq:dop_exp}
\hat{\mathbf{D}}(\mathbf{r},t) = &\hphantom{+}\sum_{\gamma} \int_0^{k_B} \Diff{k} \sqrt{\frac{\hbar \omega_k}{2}} \mathbf{D}_{\gamma,k}(\mathbf{r}) \aop{\gamma,k} \me^{-i \omega_k t}+ \sum_{\gamma'} \int_{k_B}^{\infty} \Diff{k} \sqrt{\frac{\hbar \omega_k}{2}} \mathbf{F}_{\gamma',k}(\mathbf{r}) \bop{\gamma',k}\me^{-i \omega_k t} + \text{h.c.}
\end{align}
where $\aop{\gamma,k}$ and $\bop{\gamma,k}$ are the usual bosonic annihilation operators for mode $\gamma$ and wavenumber $k$, $\mathbf{D}_{\gamma,k}$ and $\mathbf{F}_{\gamma,k}$ are mode functions, and h.c. denotes the Hermitian conjugate. We separate the expansion into low ($\aop{\gamma,k}$) and high ($\bop{\gamma,k}$) frequency bands, with $k_B$ a wavenumber between the two. For uniform waveguides, the field mode functions may be decomposed into a transverse mode function and a longitudinal plane wave,
\begin{align}
\mathbf{D}_{\gamma,k}(\mathbf{r}) = \frac{\mathbf{d}_{\gamma,k}(x,y) \me^{i k z}}{\sqrt{2\pi}}, \qquad
\mathbf{F}_{\gamma,k}(\mathbf{r}) = \frac{\mathbf{f}_{\gamma,k}(x,y) \me^{i k z}}{\sqrt{2\pi}}.
\end{align}
\indent On substituting \cref{eq:dop_exp} into \cref{eq:full_3rdo_hamil}, and considering only one mode per band, we keep only terms involving the annihilation of one photon in the high frequency band and the creation of three photons in the low frequency band, and their conjugates. We neglect the SPM and XPM terms, with the expectation that this photon source will operate in a regime where there is not enough power for these effects to be significant \cite{Helt2013}.
The interaction Hamiltonian can then be expressed as
\begin{equation}\label{eq:interaction_hamiltonian}
\begin{aligned}
\hat{H}_\text{I}(t) =& - \frac{3 \alpha^* \beta \hbar^2}{16 \pi^2 \epsilon_0} 
\int_0^{\infty} \int_0^{\infty}\int_0^{\infty}\int_0^{\infty}
\Diff{k_1}  \Diff{k_2} \Diff{k_3} \Diff{k_4} 
\sqrt{\omega_{k_1} \omega_{k_2} \omega_{k_3} \omega_{k_4}} \frac{\mathfrak{s}^*(\Delta k) \ochie}{\nbar^4}
\frac{\phi^*_s(k_4) \phi_p(k_4) \me^{-i\Delta \omega_k t}}{\mathcal{A}(k_1,k_2,k_3,k_4)} \aopd{k_1}\aopd{k_2}
+ \text{h.c.},
\end{aligned}
\end{equation}
where $\Delta k = k_4 - k_3 - k_2 - k_1$, $\Delta \omega_k = \omega_{k_4} - \omega_{k_3} - \omega_{k_2} - \omega_{k_1}$, $\left|\alpha\right|^2$ and $\left|\beta\right|^2$ are the average numbers of photons in the input classical seed and pump pulses respectively, and $\phi_s(k)$, $\phi_p(k)$ are their spectral profiles, localised in $k$. The effective mode coupling area satisfies
\vspace{-.5cm}
\begin{widetext}
	\begin{equation}\label{eq:effective_area}
	\begin{aligned}
	\frac{1}{\mathcal{A}(k_1,k_2,k_3,k_4)} \equiv& \int_{-\infty}^{\infty} \Diff{x} \int_{-\infty}^{\infty} \Diff{y} \frac{\nbar^4 \chie_{ijmn}(x,y)}{4\ochie \epsilon_0^2 n_{k_1}^2 n_{k_2}^2 n_{k_3}^2 n_{k_4}^2}\\
	&\times \left[\left(\dop{i}{k_1}\dop{j}{k_2}\dop{m}{k_3}\right)^*\fop{n}{k_4}+
	\left(\dop{i}{k_1}\dop{j}{k_2}\dop{n}{k_3}\right)^*\fop{m}{k_4}+
	\left(\dop{i}{k_1}\dop{n}{k_2}\dop{m}{k_3}\right)^*\fop{j}{k_4}+
	\left(\dop{n}{k_1}\dop{j}{k_2}\dop{m}{k_3}\right)^*\fop{i}{k_4}\right]\\
	=& \int_{-\infty}^{\infty} \Diff{x} \int_{-\infty}^{\infty} \Diff{y} \frac{\nbar^4}{4\ochie \epsilon_0^2 n_{k_1}^2 n_{k_2}^2 n_{k_3}^2 n_{k_4}^2}\\ &\times\bigg\{\left(2\chie_{1122}(x,y)+\chie_{1212}(x,y)+\chie_{1221}(x,y)\right) \left(\mathbf{d}_{k_1}\cdot\mathbf{d}_{k_2}\right)^* \left(\mathbf{d}_{k_3}\right)^*\cdot\mathbf{f}_{k_4}\\ 
	&+ \left(\chie_{1122}(x,y)+2\chie_{1212}(x,y)+\chie_{1221}(x,y)\right) \left(\mathbf{d}_{k_1}\cdot\mathbf{d}_{k_3}\right)^* \left(\mathbf{d}_{k_2}\right)^*\cdot\mathbf{f}_{k_4}\\
	&+ \left(\chie_{1122}(x,y)+\chie_{1212}(x,y)+2\chie_{1221}(x,y)\right) \left(\mathbf{d}_{k_2}\cdot\mathbf{d}_{k_3}\right)^* \left(\mathbf{d}_{k_1}\right)^*\cdot\mathbf{f}_{k_4}\bigg\},
	\end{aligned}
	\end{equation}
\end{widetext}
\vspace{-.2cm}
where $\Gamma^{(3)}_{ijmn} = \chie_{ijmn}/({\epsilon_0 n_{k_1}^2 n_{k_2}^2 n_{k_3}^2 n_{k_4}^2})$, the refractive index is abbreviated as ${n(x,y;\omega_{k_i}) = n_{k_i}}$, and the nonlinear susceptibility has been decomposed into a transverse and longitudinal part ${\chie_{ijmn}(\mathbf{r}) = \chie_{ijmn}(x,y) s(z)}$. Additionally, the typical size of a nonvanishing component of $\chie(x,y)$ is denoted $\ochie$, and $\nbar$ represents a typical value of the local refractive index, both introduced solely for convenience\cite{Yang2008}. We associate $k_1$ and $k_2$ with the generated pairs, $k_3$ with the seed and $k_4$ with the pump. Note that the second form of \cref{eq:effective_area} in vector notation holds if the material is isotropic, and both definitions for the effective area account for fields of arbitrary polarisation. The phasematching condition for SSTPDC in \cref{eq:interaction_hamiltonian} is captured in the spatial Fourier transform of the longitudinal nonlinearity profile $s(z)$:
\begin{equation}
\mathfrak{s}(k) \equiv \int_{-\infty}^{\infty} \Diff{z} s(z) \me^{-ikz}.
\end{equation}

We use \cref{eq:interaction_hamiltonian} to find the first order solution as given by \cref{eq:first_order_schro} and transform from $k$ to $\omega$. This introduces factors in the group velocity $v_g$ which account for the density of states in frequency\cite{Yang2008}.
The integration over all time yields \(\int_{-\infty}^{\infty} \me^{-i \Delta \omega_k t} = 2\pi \delta(\Delta \omega_k)\), allowing the further integration over one frequency. Now the state can be described as
\begin{equation}\label{eq:biphoton_state}
\ket{\psi_\text{out}} \approx \frac{\vac + \eta \ket{\text{II}}}{\sqrt{1+\left|\eta\right|^2}},
\end{equation}
where $\eta$ is a normalisation factor, and the biphoton state is described by 
\begin{equation}\label{eq:biphoton_state_star}
\ket{\text{II}} = \frac{1}{\sqrt{2}} \int_0^{\infty} \Diff{\omega_1} \Diff{\omega_2} \Phi(\omega_1,\omega_2) \aopd{\omega_1} \aopd{\omega_2} \vac.
\end{equation}

The joint spectral amplitude (JSA) is
\begin{multline}\label{eq:jsa}
\Phi(\omega_1,\omega_2) = \frac{1}{\eta}\Bigg(\frac{3\sqrt{2} i \alpha^* \beta \hbar}{8 \pi \epsilon_0} \int_0^{\infty} \Diff{\omega} \sqrt{\frac{\omega_1 \omega_2 \omega(\omega_1 + \omega_2 + \omega)}{v_g(\omega_1) v_g(\omega_2) v_g(\omega) v_g(\omega_1+\omega_2+\omega)}}\\
\times \frac{\mathfrak{s}^*(\Delta k)\ochie}{\nbar^4} \frac{\bar{\phi}_s^*(\omega) \bar{\phi}_p(\omega_1+\omega_2+\omega)}{\mathcal{A}\left[k(\omega_1),k(\omega_2),k(\omega),k(\omega_1+\omega_2+\omega)\right]}\Bigg).
\end{multline}
This is our main result. It fully describes the biphoton state for SSTPDC, and from \cref{eq:jsa} one can calculate the rate of photon pair production as well as arbitrary expectation values. In particular, normalising the biphoton state $\left<\text{II}|\text{II}\right>=1$ imposes the normalisation of the JSA ${\int \Diff{\omega_1}\Diff{\omega_2} \left|\Phi(\omega_1,\omega_2)\right|^2 = 1}$. From \cref{eq:biphoton_state_star}, this allows the physical interpretation of $\left|\eta\right|^2$ as the probability of pair production per pump pulse.

Approximating the seed and pump fields as Gaussians, $\bar{\phi}_j(\omega) = \left(\sqrt{\tau}_j/\pi^{1/4}\right) \me^{-\tau(\omega-\omega_j)^2/2}$, with long pulse durations $\tau_s$ and $\tau_p$ respectively, it is possible to arrive at a coarse approximation to the pair production probability
\begin{equation}\label{eq:rough_rate}
\left|\eta\right|^2 \approx \frac{4 \gamma^2 L^2}{3\pi} \sqrt{\frac{2 \tau_s^2 \tau_p^2}{\left|\beta_2(\omega_s)\right|L\left(\tau_s^2+\tau_p^2\right)}} P_p P_s,
\end{equation}
where $L$ is the interaction length, the nonlinear parameter is 
\begin{equation}\label{eq:gamma}
\gamma = \dfrac{3 \chie \omega_s}{4 \epsilon_0 \sqrt{v_g^3(\omega_s) v_g(\omega_p)} \nbar^4 \mathcal{A}},
\end{equation}
$P_s =  \hbar \omega_s \left|\alpha\right|^2/\tau_s$, and $P_p =  \hbar \omega_p \left|\beta\right|^2/\tau_p$. These are nominal average \emph{pulse} powers, related to the time averaged power by $\bar{P}_j = P_j \sigma$, where $\sigma$ is the duty cycle of a high repetition rate laser. 

As a proof-of-principle, we consider a system where we expect to be able to phasematch this process. The phasematching and efficient conversion of one-third harmonic generation (OTHG) or backwards third harmonic generation has been studied by Grubsky \etal \cite{Grubsky2005} and further refined by Zhang \etal \cite{Zhang2015a}, demonstrating how to phasematch the process in fused silica microfiber. By tuning the width of the fiber, the high frequency HE$_{21}$ mode can be phasematched with the low frequency HE$_{11}$ mode.

\begin{figure}
	\centering
	\includegraphics[width=.5\linewidth]{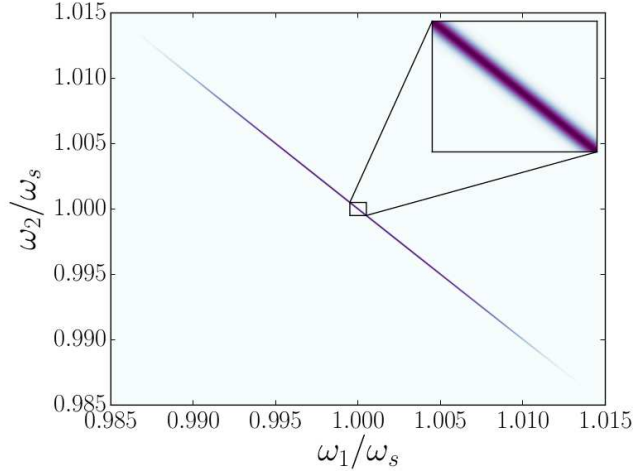}
	\caption{The normalised joint spectral intensity for SSTPDC in fused silica microfiber, with a Schmidt number of $K=106.3$, and a pair production probability of $\left|\eta\right|^2=0.028$ per pulse. Here $\omega_s/(2\pi) = \SI{187.84}{\tera\hertz}$.\label{fig:norm_jsi}}
\end{figure}

Setting the pump to the common frequency-doubled laser wavelength \SI{532}{\nano\meter}, we solve for these modes exactly and find that $n_\text{eff}^{\text{HE}_{21}}(\omega_p)=n_\text{eff}^{\text{HE}_{11}}(\omega_s)$ when the diameter of the fiber is \SI{0.790}{\micro\meter}. The effective area then is $\mathcal{A} = \SI{4.9}{\micro\meter^2}$, 
with the third order susceptibility $\chie=2.5\times10^{-22}$\si{\meter^2\per\volt^2}.
We take an interaction length $L=\SI{10}{\milli\meter}$ typical of fiber tapers.
We envisage modest pump configurations, typical of current mode-locked green sources, as shown in \cref{tab:sstpdc_values}. Note that to manage the fast walk-off, the seed pulse duration is on the order of nanoseconds. With this set of parameters, the pair production probability as given by \cref{eq:rough_rate} is $\left|\eta\right|^2=0.029$ per pulse. Without relying on these coarse approximations and assuming Gaussian pulses, the equivalent numerical result derived from \cref{eq:jsa} yields $\left|\eta\right|^2=0.028$ per pulse, across a bandwidth of \SI{3.9}{\tera\hertz}. This is sufficient for producing an effective photon source, for a pump laser with a typical 100~MHz repetition rate. The resulting joint spectral intensity (JSI) is shown in \cref{fig:norm_jsi}. It has a Schmidt number\cite{Law2004} of $K=106.3$, found from the singular value decomposition of the JSA. This indicates that the state is highly correlated, with an unheralded second order correlation function\cite{Christ2011} of  $g^{(2)}(0)=1.0094$.
\begin{table}
	\centering
	\caption{Material and pulse parameters for the SSTPDC simulation. The group velocity and group velocity dispersion were acquired from the exact solutions in microfiber.}
	
	\begin{tabular}{lcrr}
		\hline
		\textbf{Quantity} & \textbf{~~~Symbol~~~} 
		& \textbf{~~~~~Seed value $(\omega_s)$}
		& \textbf{~~~~~Pump value $(\omega_p)$}\\ \hline
		Wavelength & $\lambda$ & $\SI{1596}{\nano\meter}$ & $\SI{532}{\nano\meter}$ \\
		Refractive index & $n$ & 1.46 & 1.44 \\
		Group index & $n_g$ & 1.396 & 1.695 \\
		Group velocity dispersion & $\beta_2$ & $\SI{2344}{\pico\second^2\per\kilo\meter}$ & $\SI{-10}{\pico\second^2\per\kilo\meter}$ \\
		Pulse Power & $P$ & $\SI{1}{\watt}$ & $\SI{10}{\kilo\watt}$ \\
		Pulse duration & $\tau$ & $\SI{1.0}{\nano\second}$ & $\SI{10.0}{\pico\second}$\\
		\hline
	\end{tabular}%
	
	\label{tab:sstpdc_values}%

\end{table}%

Being a four wave mixing process with four distinct fields, the parameter space for SSTPDC is quite large, and there is room to engineer desirable JSAs. For example, decreasing the pump duration to $\tau_p = \SI{1}{\pico\second}$ decreases the Schmidt number to $K=66.9$ ($g^{(2)}(0)=1.015$), decreases the rate to $\left|\eta\right|^2 = 0.0016$ pairs per pulse, and increases the generation bandwidth to \SI{6.8}{\tera\hertz}. Doubling the interaction length $L=\SI{20}{\milli\meter}$ decreases the Schmidt number to $K=90.4$ ($g^{(2)}(0)=1.011$), decreases the generation bandwidth to \SI{2.9}{\tera\hertz}, but increases the rate to $\left|\eta\right|^2=0.076$ pairs per pulse. As a final example, a four fold increase in the group velocity dispersion decreases the rate to $\left|\eta\right|^2 = 0.014$ pairs per pulse (as we would expect from \cref{eq:rough_rate}), decreases the generation bandwidth to \SI{1.9}{\tera\hertz}, and also decreases the Schmidt number to $K=53.6$ ($g^{(2)}(0)=1.019$).
	
What about the Raman problem? As a first approximation we follow Agrawal \cite{Agrawal2006} and Lin \etal \cite{Lin2007}, working in the quasi-CW limit and expressing the Raman photon flux as ${I_u^R = \Delta \nu_u P_p L \left|g_R(\Delta)\right| [\rho+1/(\me^{h \Delta/(k_B T)}-1) ]/\mathcal{A}}$, where $u=s,i$, the detuning from the seed is $2\pi\Delta = \left|\omega - \omega_s\right|$, $\Delta \nu_u$ is a filter bandwidth. The factor $\rho=1$ for the Stokes process and is zero for the anti-Stokes process. 
The Stokes channel coincides with the idler photon band ($\omega<\omega_s$) and the anti-Stokes channel coincides with the signal photon ($\omega>\omega_s$). The Raman gain $g_R(\Delta)$ is taken directly from Agrawal \cite{Agrawal2006}. In this limit, the SFWM and SSTPDC spectral densities of the desired photon pairs are expressed as
\begin{align}
\frac{I^\text{\scriptsize{SFWM}}_u}{\Delta\nu_u} &= \left(\gamma_\text{\scriptsize{SFWM}} P_\text{sp} L\right)^2 \text{sinc}^2\left(\pi\beta_2^{\text{\scriptsize{SFWM}}}(\omega_s)L \Delta \right),\label{eq:sfwm_flux}\\
\frac{I^\text{\scriptsize{SSTPDC}}_u}{\Delta\nu_u} &= 4\left(\gamma \sqrt{P_p P_s} L\right)^2 \text{sinc}^2\left(\pi\beta_2(\omega_s)L \Delta \right),\label{eq:sstpdc_flux}
\end{align}
where $P_\text{sp}$ is the single-pump power, and similarly to \cref{eq:gamma},
\begin{equation}\label{eq:gamma_sfwm}
\gamma_\text{\scriptsize{SFWM}} = \dfrac{3 \chie \omega_s}{4 \epsilon_0 v_g^2(\omega_s) \nbar^4 \mathcal{A}_\text{\scriptsize{SFWM}}},
\end{equation}
with $\mathcal{A}_\text{\scriptsize{SFWM}}=\SI{84.0}{\micro\meter}$, $n_g^{\text{\scriptsize{SFWM}}}(\omega_s)=1.463$, and $\beta_2^\text{\scriptsize{SFWM}}(\omega_s) = -\SI{26.18}{\square\pico\second\per\kilo\meter}$. As the spontaneous Raman bandwidth is approximately \SI{40}{\tera\hertz}, and the pump and seed fields are separated by roughly \SI{400}{\tera\hertz}, we do not expect there to be any measurable contribution from the pump to the Raman noise in the pair generation band.

\begin{figure}
	\centering
	\begin{subfigure}{0.5\linewidth}
		\centering
		\includegraphics[width=.9\linewidth]{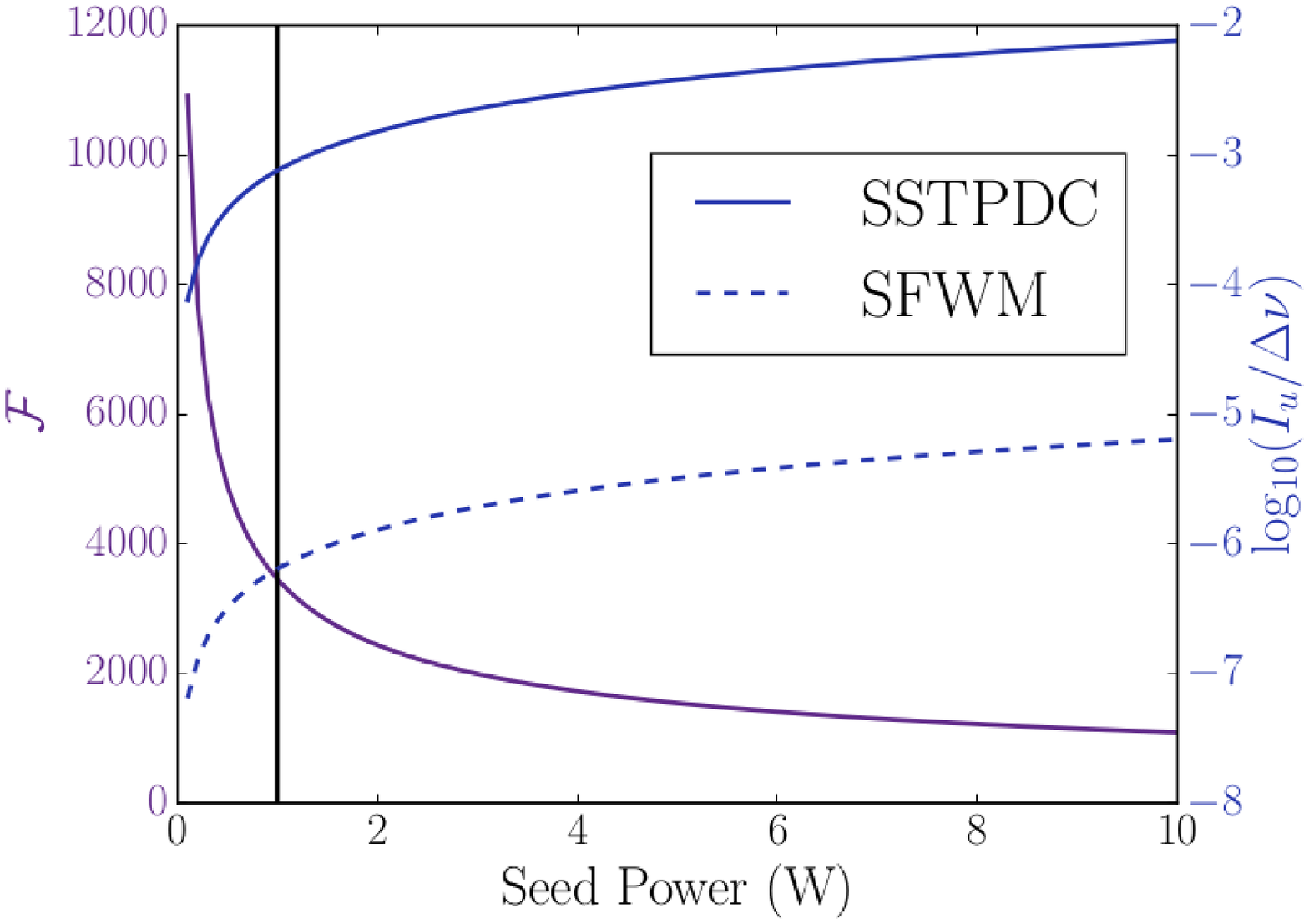}
		\caption{\label{fig:fom}}
	\end{subfigure}%
	\begin{subfigure}{0.5\linewidth}
		\centering
		\includegraphics[width=.9\linewidth]{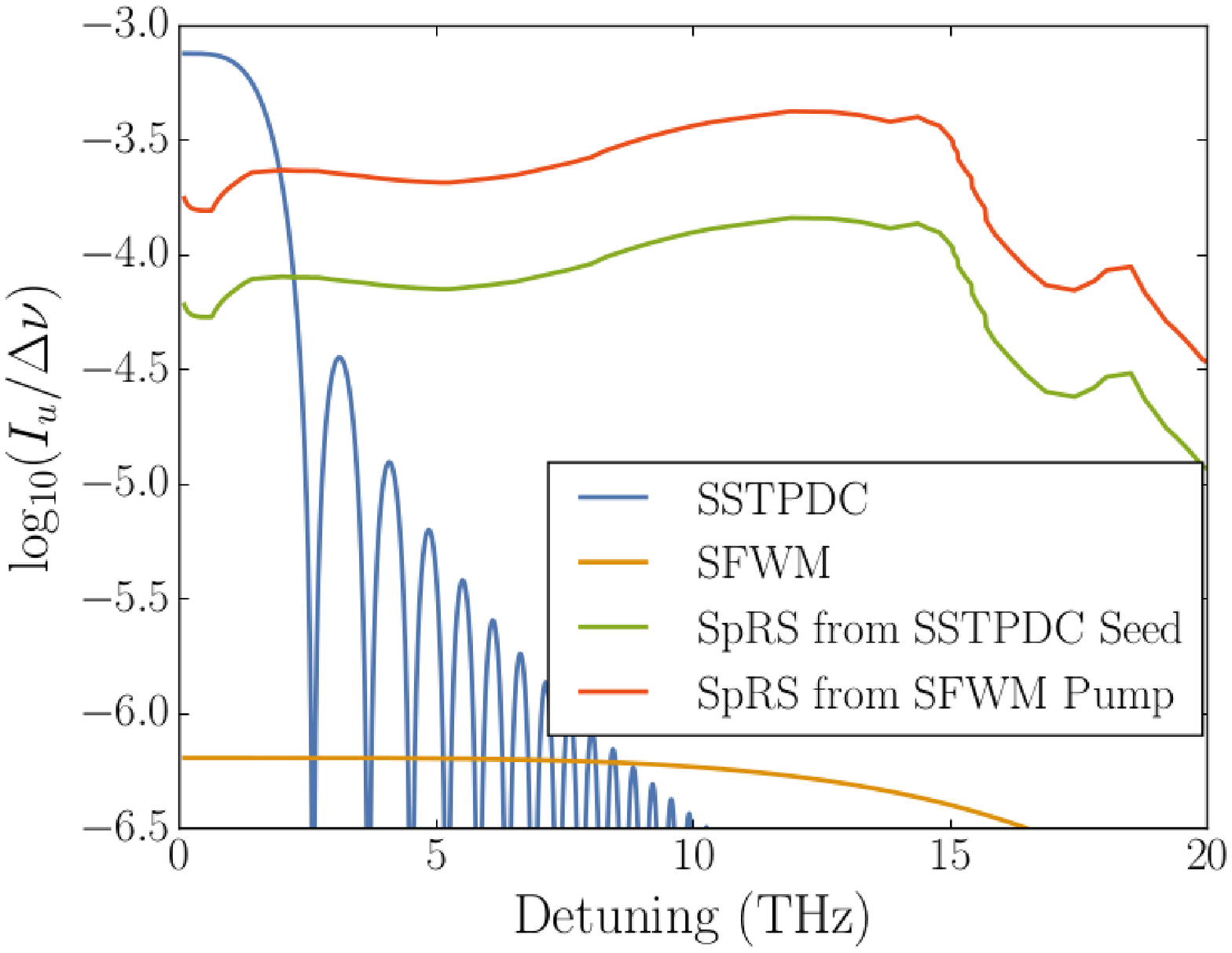}
		\caption{\label{fig:spectral_densities}}
	\end{subfigure}
	\caption{(a): The figure of merit, SSTPDC and SFWM peak spectral densities as a function of seed power. The vertical line corresponds to \SI{1}{\watt} seed power. (b): The spectral densities for single pump SFWM in standard single mode fiber, SSTPDC in microfiber, as well as their respective Raman spectral densities.}
\end{figure}

Defining the signal to noise ratio (SNR) as $\text{SNR} = I_u/I^R_u$, we define a figure of merit (FOM) to be the improvement in the SNR for SSTPDC in microfiber over the SNR of SFWM in SMF-28. Taking the single pump power to be $P_\text{sp} = \sqrt{P_s P_p}$, the FOM is expressed as
\begin{equation}\label{eq:fom}
\mathcal{F} \equiv \frac{\text{SNR}_\text{\scriptsize{SSTPDC}}}{\text{SNR}_\text{\scriptsize{SFWM}}} \approx 4\frac{\mathcal{A}_\text{\scriptsize{SFWM}}}{\mathcal{A}} \sqrt{\frac{P_p}{P_s}}.
\end{equation}
To keep $\left|\eta\right|^2 \ll 1$ (recall \cref{eq:rough_rate}), we fix $P_p$ at $\SI{10}{\kilo\watt}$ with a pulse duration of $\SI{1}{\nano\second}$ (\SI{800}{\milli\watt} average power at a repetition rate of \SI{80}{\kilo\hertz}), and plot the peak spectral densities (\cref{eq:sstpdc_flux,eq:sfwm_flux}), and FOM (\cref{eq:fom}) as functions of $P_s$. The spectral densities themselves are plotted for a representative seed power of \SI{1}{\watt} (the vertical line in \cref{fig:fom}) in \cref{fig:spectral_densities}. \Cref{fig:fom} shows a considerably improved SNR as compared with SFWM for a range of pumping configurations. \Cref{fig:spectral_densities} demonstrates a reduction in the production of Raman photons and, critically, that the SSTPDC signal lies well above the SpRS of the SSTPDC seed, whereas the SFWM signal is buried beneath the SpRS from the SFWM pump.

Evaluating the two processes for a common fiber, the ratio between $P_\text{sp}$ and $P_s$ predicts an 100-fold decrease in Raman noise. If we compare SSTPDC in microfiber with SFWM in SMF-28, taking into account that the Raman gain increases in magnitude for smaller areas, we find that the gain in microfiber is larger by a factor of $\mathcal{A}_\text{SFWM}/\mathcal{A} = 17.0$, and so the ratio that determines the suppression is approximately $P_\text{sp}\mathcal{A}/(P_s\mathcal{A}_\text{\scriptsize{SFWM}}) = 5.9$. In other systems that support third harmonic generation, for example in photonic crystal fiber\cite{Tarnowski2011}, one might expect to achieve or exceed the suppression given by $P_\text{sp}/P_s$. The freedom to separate the strong pump from the pair generation, inherent to SSTPDC, is what drives this improvement. 

We have proposed and demonstrated the feasibility of the third order photon pair generation method we call stimulated spontaneous three photon down conversion (SSTPDC). The process can be phasematched, and with realistic pump requirements produce pairs at rates sufficiently high for an effective heralded single photon source. With the strongest pump field spectrally distinct from the generated pairs, it sidesteps the issue of noise from spontaneous Raman scattering. In microfiber, we calculate an improvement in the ratio of photon pairs to uncorrelated Raman photons as compared with standard SFWM in SMF-28, for various pumping configurations.

\begin{acknowledgments}
This work was supported in part by the ARC Centre for Ultrahigh bandwidth Devices for Optical Systems (CUDOS) (Project No. CE110001018). We thank M.~J. Collins for useful discussions.
\end{acknowledgments}

\bibliography{sstpdc3}

\begin{thebibliography}{27}%
\makeatletter
\providecommand \@ifxundefined [1]{%
 \@ifx{#1\undefined}
}%
\providecommand \@ifnum [1]{%
 \ifnum #1\expandafter \@firstoftwo
 \else \expandafter \@secondoftwo
 \fi
}%
\providecommand \@ifx [1]{%
 \ifx #1\expandafter \@firstoftwo
 \else \expandafter \@secondoftwo
 \fi
}%
\providecommand \natexlab [1]{#1}%
\providecommand \enquote  [1]{``#1''}%
\providecommand \bibnamefont  [1]{#1}%
\providecommand \bibfnamefont [1]{#1}%
\providecommand \citenamefont [1]{#1}%
\providecommand \href@noop [0]{\@secondoftwo}%
\providecommand \href [0]{\begingroup \@sanitize@url \@href}%
\providecommand \@href[1]{\@@startlink{#1}\@@href}%
\providecommand \@@href[1]{\endgroup#1\@@endlink}%
\providecommand \@sanitize@url [0]{\catcode `\\12\catcode `\$12\catcode
  `\&12\catcode `\#12\catcode `\^12\catcode `\_12\catcode `\%12\relax}%
\providecommand \@@startlink[1]{}%
\providecommand \@@endlink[0]{}%
\providecommand \url  [0]{\begingroup\@sanitize@url \@url }%
\providecommand \@url [1]{\endgroup\@href {#1}{\urlprefix }}%
\providecommand \urlprefix  [0]{URL }%
\providecommand \Eprint [0]{\href }%
\providecommand \doibase [0]{http://dx.doi.org/}%
\providecommand \selectlanguage [0]{\@gobble}%
\providecommand \bibinfo  [0]{\@secondoftwo}%
\providecommand \bibfield  [0]{\@secondoftwo}%
\providecommand \translation [1]{[#1]}%
\providecommand \BibitemOpen [0]{}%
\providecommand \bibitemStop [0]{}%
\providecommand \bibitemNoStop [0]{.\EOS\space}%
\providecommand \EOS [0]{\spacefactor3000\relax}%
\providecommand \BibitemShut  [1]{\csname bibitem#1\endcsname}%
\let\auto@bib@innerbib\@empty
\bibitem [{\citenamefont {Somaschi}\ \emph {et~al.}(2016)\citenamefont
  {Somaschi}, \citenamefont {Giesz}, \citenamefont {{De Santis}}, \citenamefont
  {Loredo}, \citenamefont {Almeida}, \citenamefont {Hornecker}, \citenamefont
  {Portalupi}, \citenamefont {Grange}, \citenamefont {Anton}, \citenamefont
  {Demory}, \citenamefont {Gomez}, \citenamefont {Sagnes}, \citenamefont
  {Kimura}, \citenamefont {Lemaitre}, \citenamefont {Auffeves}, \citenamefont
  {White}, \citenamefont {Lanco},\ and\ \citenamefont
  {Senellart}}]{Somaschi2016}%
  \BibitemOpen
  \bibfield  {author} {\bibinfo {author} {\bibfnamefont {N.}~\bibnamefont
  {Somaschi}}, \bibinfo {author} {\bibfnamefont {V.}~\bibnamefont {Giesz}},
  \bibinfo {author} {\bibfnamefont {L.}~\bibnamefont {{De Santis}}}, \bibinfo
  {author} {\bibfnamefont {J.~C.}\ \bibnamefont {Loredo}}, \bibinfo {author}
  {\bibfnamefont {M.~P.}\ \bibnamefont {Almeida}}, \bibinfo {author}
  {\bibfnamefont {G.}~\bibnamefont {Hornecker}}, \bibinfo {author}
  {\bibfnamefont {S.~L.}\ \bibnamefont {Portalupi}}, \bibinfo {author}
  {\bibfnamefont {T.}~\bibnamefont {Grange}}, \bibinfo {author} {\bibfnamefont
  {C.}~\bibnamefont {Anton}}, \bibinfo {author} {\bibfnamefont
  {J.}~\bibnamefont {Demory}}, \bibinfo {author} {\bibfnamefont
  {C.}~\bibnamefont {Gomez}}, \bibinfo {author} {\bibfnamefont
  {I.}~\bibnamefont {Sagnes}}, \bibinfo {author} {\bibfnamefont {N.~D.~L.}\
  \bibnamefont {Kimura}}, \bibinfo {author} {\bibfnamefont {A.}~\bibnamefont
  {Lemaitre}}, \bibinfo {author} {\bibfnamefont {A.}~\bibnamefont {Auffeves}},
  \bibinfo {author} {\bibfnamefont {A.~G.}\ \bibnamefont {White}}, \bibinfo
  {author} {\bibfnamefont {L.}~\bibnamefont {Lanco}}, \ and\ \bibinfo {author}
  {\bibfnamefont {P.}~\bibnamefont {Senellart}},\ }\href@noop {} {\bibfield
  {journal} {\bibinfo  {journal} {Nature Photon.}\ }\textbf {\bibinfo {volume}
  {10}},\ \bibinfo {pages} {340--345} (\bibinfo {year} {2016})}\BibitemShut
  {NoStop}%
\bibitem [{\citenamefont {Darqui\'{e}}\ \emph {et~al.}(2005)\citenamefont
  {Darqui\'{e}}, \citenamefont {Jones}, \citenamefont {Dingjan}, \citenamefont
  {Beugnon}, \citenamefont {Bergamini}, \citenamefont {Sortais}, \citenamefont
  {Messin}, \citenamefont {Browaeys},\ and\ \citenamefont
  {Grangier}}]{Darquie2005}%
  \BibitemOpen
  \bibfield  {author} {\bibinfo {author} {\bibfnamefont {B.}~\bibnamefont
  {Darqui\'{e}}}, \bibinfo {author} {\bibfnamefont {M.}~\bibnamefont {Jones}},
  \bibinfo {author} {\bibfnamefont {J.}~\bibnamefont {Dingjan}}, \bibinfo
  {author} {\bibfnamefont {J.}~\bibnamefont {Beugnon}}, \bibinfo {author}
  {\bibfnamefont {S.}~\bibnamefont {Bergamini}}, \bibinfo {author}
  {\bibfnamefont {Y.}~\bibnamefont {Sortais}}, \bibinfo {author} {\bibfnamefont
  {G.}~\bibnamefont {Messin}}, \bibinfo {author} {\bibfnamefont
  {A.}~\bibnamefont {Browaeys}}, \ and\ \bibinfo {author} {\bibfnamefont
  {P.}~\bibnamefont {Grangier}},\ }\href@noop {} {\bibfield  {journal}
  {\bibinfo  {journal} {Science}\ }\textbf {\bibinfo {volume} {309}},\ \bibinfo
  {pages} {454--456} (\bibinfo {year} {2005})}\BibitemShut {NoStop}%
\bibitem [{\citenamefont {Hennessy}\ \emph {et~al.}(2007)\citenamefont
  {Hennessy}, \citenamefont {Badolato}, \citenamefont {Winger}, \citenamefont
  {Gerace}, \citenamefont {Atat\"{u}re}, \citenamefont {Gulde}, \citenamefont
  {F\"{a}lt}, \citenamefont {Hu},\ and\ \citenamefont
  {Imamo\u{g}lu}}]{Hennessy2007}%
  \BibitemOpen
  \bibfield  {author} {\bibinfo {author} {\bibfnamefont {K.}~\bibnamefont
  {Hennessy}}, \bibinfo {author} {\bibfnamefont {A.}~\bibnamefont {Badolato}},
  \bibinfo {author} {\bibfnamefont {M.}~\bibnamefont {Winger}}, \bibinfo
  {author} {\bibfnamefont {D.}~\bibnamefont {Gerace}}, \bibinfo {author}
  {\bibfnamefont {M.}~\bibnamefont {Atat\"{u}re}}, \bibinfo {author}
  {\bibfnamefont {S.}~\bibnamefont {Gulde}}, \bibinfo {author} {\bibfnamefont
  {S.}~\bibnamefont {F\"{a}lt}}, \bibinfo {author} {\bibfnamefont {E.~L.}\
  \bibnamefont {Hu}}, \ and\ \bibinfo {author} {\bibfnamefont {A.}~\bibnamefont
  {Imamo\u{g}lu}},\ }\href
  {http://www.nature.com/doifinder/10.1038/nature05586} {\bibfield  {journal}
  {\bibinfo  {journal} {Nature}\ }\textbf {\bibinfo {volume} {445}},\ \bibinfo
  {pages} {896--899} (\bibinfo {year} {2007})}\BibitemShut {NoStop}%
\bibitem [{\citenamefont {Albrecht}\ \emph {et~al.}(2014)\citenamefont
  {Albrecht}, \citenamefont {Bommer}, \citenamefont {Pauly}, \citenamefont
  {M\"{u}cklich}, \citenamefont {Schell}, \citenamefont {Engel}, \citenamefont
  {Schr\"{o}der}, \citenamefont {Benson}, \citenamefont {Reichel},\ and\
  \citenamefont {Becher}}]{Albrecht2014}%
  \BibitemOpen
  \bibfield  {author} {\bibinfo {author} {\bibfnamefont {R.}~\bibnamefont
  {Albrecht}}, \bibinfo {author} {\bibfnamefont {A.}~\bibnamefont {Bommer}},
  \bibinfo {author} {\bibfnamefont {C.}~\bibnamefont {Pauly}}, \bibinfo
  {author} {\bibfnamefont {F.}~\bibnamefont {M\"{u}cklich}}, \bibinfo {author}
  {\bibfnamefont {A.~W.}\ \bibnamefont {Schell}}, \bibinfo {author}
  {\bibfnamefont {P.}~\bibnamefont {Engel}}, \bibinfo {author} {\bibfnamefont
  {T.}~\bibnamefont {Schr\"{o}der}}, \bibinfo {author} {\bibfnamefont
  {O.}~\bibnamefont {Benson}}, \bibinfo {author} {\bibfnamefont
  {J.}~\bibnamefont {Reichel}}, \ and\ \bibinfo {author} {\bibfnamefont
  {C.}~\bibnamefont {Becher}},\ }\href
  {http://scitation.aip.org/content/aip/journal/apl/105/7/10.1063/1.4893612}
  {\bibfield  {journal} {\bibinfo  {journal} {Appl. Phys. Lett.}\ }\textbf
  {\bibinfo {volume} {105}},\ \bibinfo {pages} {073113} (\bibinfo {year}
  {2014})}\BibitemShut {NoStop}%
\bibitem [{\citenamefont {Kwiat}\ \emph {et~al.}(1995)\citenamefont {Kwiat},
  \citenamefont {Mattle}, \citenamefont {Weinfurter},\ and\ \citenamefont
  {Zeilinger}}]{Kwiat1995}%
  \BibitemOpen
  \bibfield  {author} {\bibinfo {author} {\bibfnamefont {P.~G.}\ \bibnamefont
  {Kwiat}}, \bibinfo {author} {\bibfnamefont {K.}~\bibnamefont {Mattle}},
  \bibinfo {author} {\bibfnamefont {H.}~\bibnamefont {Weinfurter}}, \ and\
  \bibinfo {author} {\bibfnamefont {A.}~\bibnamefont {Zeilinger}},\ }\href@noop
  {} {\bibfield  {journal} {\bibinfo  {journal} {Phys. Rev. Lett.}\ }\textbf
  {\bibinfo {volume} {75}},\ \bibinfo {pages} {4337--4342} (\bibinfo {year}
  {1995})}\BibitemShut {NoStop}%
\bibitem [{\citenamefont {Bonfrate}\ \emph {et~al.}(1999)\citenamefont
  {Bonfrate}, \citenamefont {Pruneri}, \citenamefont {Kazansky}, \citenamefont
  {Tapster},\ and\ \citenamefont {Rarity}}]{Bonfrate1999}%
  \BibitemOpen
  \bibfield  {author} {\bibinfo {author} {\bibfnamefont {G.}~\bibnamefont
  {Bonfrate}}, \bibinfo {author} {\bibfnamefont {V.}~\bibnamefont {Pruneri}},
  \bibinfo {author} {\bibfnamefont {P.~G.}\ \bibnamefont {Kazansky}}, \bibinfo
  {author} {\bibfnamefont {P.}~\bibnamefont {Tapster}}, \ and\ \bibinfo
  {author} {\bibfnamefont {J.~G.}\ \bibnamefont {Rarity}},\ }\href
  {http://eprints.soton.ac.uk/77806/$\backslash$nhttp://scitation.aip.org/content/aip/journal/apl/75/16/10.1063/1.125013}
  {\bibfield  {journal} {\bibinfo  {journal} {Appl. Phys. Lett.}\ }\textbf
  {\bibinfo {volume} {75}},\ \bibinfo {pages} {2356} (\bibinfo {year}
  {1999})}\BibitemShut {NoStop}%
\bibitem [{\citenamefont {Tanzilli}\ \emph {et~al.}(2001)\citenamefont
  {Tanzilli}, \citenamefont {{De Riedmatten}}, \citenamefont {Tittel},
  \citenamefont {Zbinden}, \citenamefont {Baldi}, \citenamefont {{De Macheli}},
  \citenamefont {Ostrowsky},\ and\ \citenamefont {Gisin}}]{Tanzilli2001}%
  \BibitemOpen
  \bibfield  {author} {\bibinfo {author} {\bibfnamefont {S.}~\bibnamefont
  {Tanzilli}}, \bibinfo {author} {\bibfnamefont {H.}~\bibnamefont {{De
  Riedmatten}}}, \bibinfo {author} {\bibfnamefont {W.}~\bibnamefont {Tittel}},
  \bibinfo {author} {\bibfnamefont {H.}~\bibnamefont {Zbinden}}, \bibinfo
  {author} {\bibfnamefont {P.}~\bibnamefont {Baldi}}, \bibinfo {author}
  {\bibfnamefont {M.}~\bibnamefont {{De Macheli}}}, \bibinfo {author}
  {\bibfnamefont {D.~B.}\ \bibnamefont {Ostrowsky}}, \ and\ \bibinfo {author}
  {\bibfnamefont {N.}~\bibnamefont {Gisin}},\ }\href@noop {} {\bibfield
  {journal} {\bibinfo  {journal} {Electron. Lett.}\ }\textbf {\bibinfo {volume}
  {37}},\ \bibinfo {pages} {26--28} (\bibinfo {year} {2001})}\BibitemShut
  {NoStop}%
\bibitem [{\citenamefont {Fiorentino}\ \emph {et~al.}(2002)\citenamefont
  {Fiorentino}, \citenamefont {Voss}, \citenamefont {Sharping},\ and\
  \citenamefont {Kumar}}]{Fiorentino2002}%
  \BibitemOpen
  \bibfield  {author} {\bibinfo {author} {\bibfnamefont {M.}~\bibnamefont
  {Fiorentino}}, \bibinfo {author} {\bibfnamefont {P.~L.}\ \bibnamefont
  {Voss}}, \bibinfo {author} {\bibfnamefont {J.~E.}\ \bibnamefont {Sharping}},
  \ and\ \bibinfo {author} {\bibfnamefont {P.}~\bibnamefont {Kumar}},\
  }\href@noop {} {\bibfield  {journal} {\bibinfo  {journal} {IEEE Photonics
  Tech. Lett.}\ }\textbf {\bibinfo {volume} {14}},\ \bibinfo {pages} {983}
  (\bibinfo {year} {2002})}\BibitemShut {NoStop}%
\bibitem [{\citenamefont {Inoue}\ and\ \citenamefont
  {Shimizu}(2004)}]{Inoue2004}%
  \BibitemOpen
  \bibfield  {author} {\bibinfo {author} {\bibfnamefont {K.}~\bibnamefont
  {Inoue}}\ and\ \bibinfo {author} {\bibfnamefont {K.}~\bibnamefont
  {Shimizu}},\ }\href@noop {} {\bibfield  {journal} {\bibinfo  {journal} {Jpn.
  J. Appl. Phys. 1}\ }\textbf {\bibinfo {volume} {43}},\ \bibinfo {pages}
  {8048--8052} (\bibinfo {year} {2004})}\BibitemShut {NoStop}%
\bibitem [{\citenamefont {Sharping}\ \emph {et~al.}(2006)\citenamefont
  {Sharping}, \citenamefont {Lee}, \citenamefont {Foster}, \citenamefont
  {Turner}, \citenamefont {Schmidt}, \citenamefont {Lipson}, \citenamefont
  {Gaeta},\ and\ \citenamefont {Kumar}}]{Sharping2006}%
  \BibitemOpen
  \bibfield  {author} {\bibinfo {author} {\bibfnamefont {J.~E.}\ \bibnamefont
  {Sharping}}, \bibinfo {author} {\bibfnamefont {K.~F.}\ \bibnamefont {Lee}},
  \bibinfo {author} {\bibfnamefont {M.~A.}\ \bibnamefont {Foster}}, \bibinfo
  {author} {\bibfnamefont {A.~C.}\ \bibnamefont {Turner}}, \bibinfo {author}
  {\bibfnamefont {B.~S.}\ \bibnamefont {Schmidt}}, \bibinfo {author}
  {\bibfnamefont {M.}~\bibnamefont {Lipson}}, \bibinfo {author} {\bibfnamefont
  {A.~L.}\ \bibnamefont {Gaeta}}, \ and\ \bibinfo {author} {\bibfnamefont
  {P.}~\bibnamefont {Kumar}},\ }\href@noop {} {\bibfield  {journal} {\bibinfo
  {journal} {Opt. Express}\ }\textbf {\bibinfo {volume} {14}},\ \bibinfo
  {pages} {12388--12393} (\bibinfo {year} {2006})}\BibitemShut {NoStop}%
\bibitem [{\citenamefont {Lin}, \citenamefont {Yaman},\ and\ \citenamefont
  {Agrawal}(2007)}]{Lin2007}%
  \BibitemOpen
  \bibfield  {author} {\bibinfo {author} {\bibfnamefont {Q.}~\bibnamefont
  {Lin}}, \bibinfo {author} {\bibfnamefont {F.}~\bibnamefont {Yaman}}, \ and\
  \bibinfo {author} {\bibfnamefont {G.~P.}\ \bibnamefont {Agrawal}},\
  }\href@noop {} {\bibfield  {journal} {\bibinfo  {journal} {Phys. Rev. A}\
  }\textbf {\bibinfo {volume} {75}},\ \bibinfo {pages} {023803} (\bibinfo
  {year} {2007})}\BibitemShut {NoStop}%
\bibitem [{\citenamefont {Collins}\ \emph {et~al.}(2012)\citenamefont
  {Collins}, \citenamefont {Clark}, \citenamefont {He}, \citenamefont {Choi},
  \citenamefont {Williams}, \citenamefont {Judge}, \citenamefont {Madden},
  \citenamefont {Withford}, \citenamefont {Steel}, \citenamefont
  {Luther-Davies}, \citenamefont {Xiong},\ and\ \citenamefont
  {Eggleton}}]{Collins2012a}%
  \BibitemOpen
  \bibfield  {author} {\bibinfo {author} {\bibfnamefont {M.~J.}\ \bibnamefont
  {Collins}}, \bibinfo {author} {\bibfnamefont {A.~S.}\ \bibnamefont {Clark}},
  \bibinfo {author} {\bibfnamefont {J.}~\bibnamefont {He}}, \bibinfo {author}
  {\bibfnamefont {D.-Y.}\ \bibnamefont {Choi}}, \bibinfo {author}
  {\bibfnamefont {R.~J.}\ \bibnamefont {Williams}}, \bibinfo {author}
  {\bibfnamefont {A.~C.}\ \bibnamefont {Judge}}, \bibinfo {author}
  {\bibfnamefont {S.~J.}\ \bibnamefont {Madden}}, \bibinfo {author}
  {\bibfnamefont {M.~J.}\ \bibnamefont {Withford}}, \bibinfo {author}
  {\bibfnamefont {M.~J.}\ \bibnamefont {Steel}}, \bibinfo {author}
  {\bibfnamefont {B.}~\bibnamefont {Luther-Davies}}, \bibinfo {author}
  {\bibfnamefont {C.}~\bibnamefont {Xiong}}, \ and\ \bibinfo {author}
  {\bibfnamefont {B.~J.}\ \bibnamefont {Eggleton}},\ }\href@noop {} {\bibfield
  {journal} {\bibinfo  {journal} {Opt. Lett.}\ }\textbf {\bibinfo {volume}
  {37}},\ \bibinfo {pages} {3393} (\bibinfo {year} {2012})}\BibitemShut
  {NoStop}%
\bibitem [{\citenamefont {Clark}\ \emph {et~al.}(2011)\citenamefont {Clark},
  \citenamefont {Bell}, \citenamefont {Fulconis}, \citenamefont {Halder},
  \citenamefont {Cemlyn}, \citenamefont {Alibart}, \citenamefont {Xiong},
  \citenamefont {Wadsworth},\ and\ \citenamefont {Rarity}}]{Clark2011}%
  \BibitemOpen
  \bibfield  {author} {\bibinfo {author} {\bibfnamefont {A.}~\bibnamefont
  {Clark}}, \bibinfo {author} {\bibfnamefont {B.}~\bibnamefont {Bell}},
  \bibinfo {author} {\bibfnamefont {J.}~\bibnamefont {Fulconis}}, \bibinfo
  {author} {\bibfnamefont {M.~M.}\ \bibnamefont {Halder}}, \bibinfo {author}
  {\bibfnamefont {B.}~\bibnamefont {Cemlyn}}, \bibinfo {author} {\bibfnamefont
  {O.}~\bibnamefont {Alibart}}, \bibinfo {author} {\bibfnamefont
  {C.}~\bibnamefont {Xiong}}, \bibinfo {author} {\bibfnamefont {W.~J.}\
  \bibnamefont {Wadsworth}}, \ and\ \bibinfo {author} {\bibfnamefont {J.~G.}\
  \bibnamefont {Rarity}},\ }\href@noop {} {\bibfield  {journal} {\bibinfo
  {journal} {New J. Phys.}\ }\textbf {\bibinfo {volume} {13}},\ \bibinfo
  {pages} {065009} (\bibinfo {year} {2011})}\BibitemShut {NoStop}%
\bibitem [{\citenamefont {Li}\ \emph {et~al.}(2006)\citenamefont {Li},
  \citenamefont {Liang}, \citenamefont {{Fook Lee}}, \citenamefont {Chen},
  \citenamefont {Voss},\ and\ \citenamefont {Kumar}}]{Li2006}%
  \BibitemOpen
  \bibfield  {author} {\bibinfo {author} {\bibfnamefont {X.}~\bibnamefont
  {Li}}, \bibinfo {author} {\bibfnamefont {C.}~\bibnamefont {Liang}}, \bibinfo
  {author} {\bibfnamefont {K.}~\bibnamefont {{Fook Lee}}}, \bibinfo {author}
  {\bibfnamefont {J.}~\bibnamefont {Chen}}, \bibinfo {author} {\bibfnamefont
  {P.~L.}\ \bibnamefont {Voss}}, \ and\ \bibinfo {author} {\bibfnamefont
  {P.}~\bibnamefont {Kumar}},\ }\href@noop {} {\bibfield  {journal} {\bibinfo
  {journal} {Phys. Rev. A}\ }\textbf {\bibinfo {volume} {73}},\ \bibinfo
  {pages} {534--544} (\bibinfo {year} {2006})}\BibitemShut {NoStop}%
\bibitem [{\citenamefont {He}\ \emph {et~al.}(2012)\citenamefont {He},
  \citenamefont {Xiong}, \citenamefont {Clark}, \citenamefont {Collins},
  \citenamefont {Gai}, \citenamefont {Choi}, \citenamefont {Madden},
  \citenamefont {Luther-Davies},\ and\ \citenamefont {Eggleton}}]{He2012}%
  \BibitemOpen
  \bibfield  {author} {\bibinfo {author} {\bibfnamefont {J.}~\bibnamefont
  {He}}, \bibinfo {author} {\bibfnamefont {C.}~\bibnamefont {Xiong}}, \bibinfo
  {author} {\bibfnamefont {A.~S.}\ \bibnamefont {Clark}}, \bibinfo {author}
  {\bibfnamefont {M.~J.}\ \bibnamefont {Collins}}, \bibinfo {author}
  {\bibfnamefont {X.}~\bibnamefont {Gai}}, \bibinfo {author} {\bibfnamefont
  {D.~Y.}\ \bibnamefont {Choi}}, \bibinfo {author} {\bibfnamefont {S.~J.}\
  \bibnamefont {Madden}}, \bibinfo {author} {\bibfnamefont {B.}~\bibnamefont
  {Luther-Davies}}, \ and\ \bibinfo {author} {\bibfnamefont {B.~J.}\
  \bibnamefont {Eggleton}},\ }\href@noop {} {\bibfield  {journal} {\bibinfo
  {journal} {J. Appl. Phys.}\ }\textbf {\bibinfo {volume} {112}},\ \bibinfo
  {pages} {9--14} (\bibinfo {year} {2012})}\BibitemShut {NoStop}%
\bibitem [{\citenamefont {Bencheikh}\ \emph {et~al.}(2007)\citenamefont
  {Bencheikh}, \citenamefont {Gravier}, \citenamefont {Douady}, \citenamefont
  {Levenson},\ and\ \citenamefont {Boulanger}}]{Bencheikh2007}%
  \BibitemOpen
  \bibfield  {author} {\bibinfo {author} {\bibfnamefont {K.}~\bibnamefont
  {Bencheikh}}, \bibinfo {author} {\bibfnamefont {F.}~\bibnamefont {Gravier}},
  \bibinfo {author} {\bibfnamefont {J.}~\bibnamefont {Douady}}, \bibinfo
  {author} {\bibfnamefont {A.}~\bibnamefont {Levenson}}, \ and\ \bibinfo
  {author} {\bibfnamefont {B.}~\bibnamefont {Boulanger}},\ }\href@noop {}
  {\bibfield  {journal} {\bibinfo  {journal} {C. R. Phys.}\ }\textbf {\bibinfo
  {volume} {8}},\ \bibinfo {pages} {206--220} (\bibinfo {year}
  {2007})}\BibitemShut {NoStop}%
\bibitem [{\citenamefont {H{\"{u}}bel}\ \emph {et~al.}(2010)\citenamefont
  {H{\"{u}}bel}, \citenamefont {Hamel}, \citenamefont {Fedrizzi}, \citenamefont
  {Ramelow}, \citenamefont {Resch},\ and\ \citenamefont
  {Jennewein}}]{Hubel2010}%
  \BibitemOpen
  \bibfield  {author} {\bibinfo {author} {\bibfnamefont {H.}~\bibnamefont
  {H{\"{u}}bel}}, \bibinfo {author} {\bibfnamefont {D.~R.}\ \bibnamefont
  {Hamel}}, \bibinfo {author} {\bibfnamefont {A.}~\bibnamefont {Fedrizzi}},
  \bibinfo {author} {\bibfnamefont {S.}~\bibnamefont {Ramelow}}, \bibinfo
  {author} {\bibfnamefont {K.~J.}\ \bibnamefont {Resch}}, \ and\ \bibinfo
  {author} {\bibfnamefont {T.}~\bibnamefont {Jennewein}},\ }\href
  {http://www.nature.com/nature/journal/v466/n7306/full/nature09175.html}
  {\bibfield  {journal} {\bibinfo  {journal} {Nature}\ }\textbf {\bibinfo
  {volume} {466}},\ \bibinfo {pages} {601--3} (\bibinfo {year}
  {2010})}\BibitemShut {NoStop}%
\bibitem [{\citenamefont {Dot}\ \emph {et~al.}(2012)\citenamefont {Dot},
  \citenamefont {Borne}, \citenamefont {Boulanger}, \citenamefont {Bencheikh},\
  and\ \citenamefont {Levenson}}]{Dot2012}%
  \BibitemOpen
  \bibfield  {author} {\bibinfo {author} {\bibfnamefont {A.}~\bibnamefont
  {Dot}}, \bibinfo {author} {\bibfnamefont {A.}~\bibnamefont {Borne}}, \bibinfo
  {author} {\bibfnamefont {B.}~\bibnamefont {Boulanger}}, \bibinfo {author}
  {\bibfnamefont {K.}~\bibnamefont {Bencheikh}}, \ and\ \bibinfo {author}
  {\bibfnamefont {J.~A.}\ \bibnamefont {Levenson}},\ }\href@noop {} {\bibfield
  {journal} {\bibinfo  {journal} {Phys. Rev. A}\ }\textbf {\bibinfo {volume}
  {85}},\ \bibinfo {pages} {023809} (\bibinfo {year} {2012})}\BibitemShut
  {NoStop}%
\bibitem [{\citenamefont {Hamel}\ \emph {et~al.}(2014)\citenamefont {Hamel},
  \citenamefont {Shalm}, \citenamefont {H{\"{u}}bel}, \citenamefont {Miller},
  \citenamefont {Marsili}, \citenamefont {Verma}, \citenamefont {Mirin},
  \citenamefont {Nam}, \citenamefont {Resch},\ and\ \citenamefont
  {Jennewein}}]{Hamel2014}%
  \BibitemOpen
  \bibfield  {author} {\bibinfo {author} {\bibfnamefont {D.~R.}\ \bibnamefont
  {Hamel}}, \bibinfo {author} {\bibfnamefont {L.~K.}\ \bibnamefont {Shalm}},
  \bibinfo {author} {\bibfnamefont {H.}~\bibnamefont {H{\"{u}}bel}}, \bibinfo
  {author} {\bibfnamefont {A.~J.}\ \bibnamefont {Miller}}, \bibinfo {author}
  {\bibfnamefont {F.}~\bibnamefont {Marsili}}, \bibinfo {author} {\bibfnamefont
  {V.~B.}\ \bibnamefont {Verma}}, \bibinfo {author} {\bibfnamefont {R.~P.}\
  \bibnamefont {Mirin}}, \bibinfo {author} {\bibfnamefont {S.~W.}\ \bibnamefont
  {Nam}}, \bibinfo {author} {\bibfnamefont {K.~J.}\ \bibnamefont {Resch}}, \
  and\ \bibinfo {author} {\bibfnamefont {T.}~\bibnamefont {Jennewein}},\ }\href
  {http://dx.doi.org/10.1038/nphoton.2014.218} {\bibfield  {journal} {\bibinfo
  {journal} {Nature Photon.}\ }\textbf {\bibinfo {volume} {8}},\ \bibinfo
  {pages} {801--807} (\bibinfo {year} {2014})}\BibitemShut {NoStop}%
\bibitem [{\citenamefont {Yang}, \citenamefont {Liscidini},\ and\ \citenamefont
  {Sipe}(2008)}]{Yang2008}%
  \BibitemOpen
  \bibfield  {author} {\bibinfo {author} {\bibfnamefont {Z.}~\bibnamefont
  {Yang}}, \bibinfo {author} {\bibfnamefont {M.}~\bibnamefont {Liscidini}}, \
  and\ \bibinfo {author} {\bibfnamefont {J.}~\bibnamefont {Sipe}},\ }\href
  {http://link.aps.org/doi/10.1103/PhysRevA.77.033808} {\bibfield  {journal}
  {\bibinfo  {journal} {Phys. Rev. A}\ }\textbf {\bibinfo {volume} {77}},\
  \bibinfo {pages} {033808} (\bibinfo {year} {2008})}\BibitemShut {NoStop}%
\bibitem [{\citenamefont {Helt}, \citenamefont {Steel},\ and\ \citenamefont
  {Sipe}(2013)}]{Helt2013}%
  \BibitemOpen
  \bibfield  {author} {\bibinfo {author} {\bibfnamefont {L.~G.}\ \bibnamefont
  {Helt}}, \bibinfo {author} {\bibfnamefont {M.~J.}\ \bibnamefont {Steel}}, \
  and\ \bibinfo {author} {\bibfnamefont {J.~E.}\ \bibnamefont {Sipe}},\ }\href
  {http://scitation.aip.org/content/aip/journal/apl/102/20/10.1063/1.4807503}
  {\bibfield  {journal} {\bibinfo  {journal} {Appl. Phys. Lett.}\ }\textbf
  {\bibinfo {volume} {102}},\ \bibinfo {pages} {201106} (\bibinfo {year}
  {2013})}\BibitemShut {NoStop}%
\bibitem [{\citenamefont {Grubsky}\ and\ \citenamefont
  {Savchenko}(2005)}]{Grubsky2005}%
  \BibitemOpen
  \bibfield  {author} {\bibinfo {author} {\bibfnamefont {V.}~\bibnamefont
  {Grubsky}}\ and\ \bibinfo {author} {\bibfnamefont {A.}~\bibnamefont
  {Savchenko}},\ }\href@noop {} {\bibfield  {journal} {\bibinfo  {journal}
  {Opt. Express}\ }\textbf {\bibinfo {volume} {13}},\ \bibinfo {pages}
  {6798--6806} (\bibinfo {year} {2005})}\BibitemShut {NoStop}%
\bibitem [{\citenamefont {Zhang}, \citenamefont {Sun},\ and\ \citenamefont
  {Song}(2015)}]{Zhang2015a}%
  \BibitemOpen
  \bibfield  {author} {\bibinfo {author} {\bibfnamefont {J.}~\bibnamefont
  {Zhang}}, \bibinfo {author} {\bibfnamefont {Y.}~\bibnamefont {Sun}}, \ and\
  \bibinfo {author} {\bibfnamefont {Q.}~\bibnamefont {Song}},\ }\href
  {https://www.osapublishing.org/abstract.cfm?URI=oe-23-13-17407} {\bibfield
  {journal} {\bibinfo  {journal} {Opt. Express}\ }\textbf {\bibinfo {volume}
  {23}},\ \bibinfo {pages} {17407} (\bibinfo {year} {2015})}\BibitemShut
  {NoStop}%
\bibitem [{\citenamefont {Law}\ and\ \citenamefont {Eberly}(2004)}]{Law2004}%
  \BibitemOpen
  \bibfield  {author} {\bibinfo {author} {\bibfnamefont {C.~K.}\ \bibnamefont
  {Law}}\ and\ \bibinfo {author} {\bibfnamefont {J.~H.}\ \bibnamefont
  {Eberly}},\ }\href@noop {} {\bibfield  {journal} {\bibinfo  {journal} {Phys.
  Rev. Lett.}\ }\textbf {\bibinfo {volume} {92}},\ \bibinfo {pages} {127903}
  (\bibinfo {year} {2004})}\BibitemShut {NoStop}%
\bibitem [{\citenamefont {Christ}\ \emph {et~al.}(2011)\citenamefont {Christ},
  \citenamefont {Laiho}, \citenamefont {Eckstein}, \citenamefont {Cassemiro},\
  and\ \citenamefont {Silberhorn}}]{Christ2011}%
  \BibitemOpen
  \bibfield  {author} {\bibinfo {author} {\bibfnamefont {A.}~\bibnamefont
  {Christ}}, \bibinfo {author} {\bibfnamefont {K.}~\bibnamefont {Laiho}},
  \bibinfo {author} {\bibfnamefont {A.}~\bibnamefont {Eckstein}}, \bibinfo
  {author} {\bibfnamefont {K.~N.}\ \bibnamefont {Cassemiro}}, \ and\ \bibinfo
  {author} {\bibfnamefont {C.}~\bibnamefont {Silberhorn}},\ }\href@noop {}
  {\bibfield  {journal} {\bibinfo  {journal} {New J. Phys.}\ }\textbf {\bibinfo
  {volume} {13}},\ \bibinfo {pages} {033027} (\bibinfo {year}
  {2011})}\BibitemShut {NoStop}%
\bibitem [{\citenamefont {Agrawal}(2006)}]{Agrawal2006}%
  \BibitemOpen
  \bibfield  {author} {\bibinfo {author} {\bibfnamefont {G.~P.}\ \bibnamefont
  {Agrawal}},\ }\href@noop {} {\emph {\bibinfo {title} {Nonlinear Fiber
  Optics}}},\ \bibinfo {edition} {4th}\ ed.\ (\bibinfo  {publisher}
  {Elsevier},\ \bibinfo {year} {2006})\BibitemShut {NoStop}%
\bibitem [{\citenamefont {Tarnowski}, \citenamefont {Kibler},\ and\
  \citenamefont {Urbanczyk}(2011)}]{Tarnowski2011}%
  \BibitemOpen
  \bibfield  {author} {\bibinfo {author} {\bibfnamefont {K.}~\bibnamefont
  {Tarnowski}}, \bibinfo {author} {\bibfnamefont {B.}~\bibnamefont {Kibler}}, \
  and\ \bibinfo {author} {\bibfnamefont {W.}~\bibnamefont {Urbanczyk}},\ }\href
  {http://www.opticsinfobase.org/abstract.cfm?URI=josab-28-9-2075} {\bibfield
  {journal} {\bibinfo  {journal} {J. Opt. Soc. Am. B}\ }\textbf {\bibinfo
  {volume} {28}},\ \bibinfo {pages} {2075} (\bibinfo {year}
  {2011})}\BibitemShut {NoStop}%
\end{thebibliography}%

\end{document}